\documentclass[aps,preprint,nofootinbib,superscriptaddress]{revtex4-1}
\usepackage{amssymb}

\usepackage{epsfig}
\usepackage[bookmarksnumbered,bookmarksopen,colorlinks,citecolor=blue,linkcolor=blue]{hyperref}

\usepackage{amsmath}

\begin{document}


\title{Properties of nuclear matter from macroscopic-microscopic mass formulas }

\author{Ning Wang}
\email{wangning@gxnu.edu.cn}\affiliation{ Department of Physics,
Guangxi Normal University, Guilin 541004, China }
\affiliation{ State Key Laboratory of Theoretical Physics, Institute of Theoretical Physics, Chinese Academy of Sciences, Beijing 100190, China}

\author{Min Liu}
\affiliation{ Department of Physics,
Guangxi Normal University, Guilin 541004, China }

\author{Li Ou}
\affiliation{ Department of Physics,
Guangxi Normal University, Guilin 541004, China }

\author{Yingxun Zhang}
\affiliation{China Institute of Atomic Energy, Beijing 102413, China}

\begin{abstract}
Based on the standard Skyrme energy density functionals together with the extended Thomas-Fermi approach, the properties of symmetric and asymmetric nuclear matter represented in two macroscopic-microscopic mass formulas: Lublin-Strasbourg nuclear drop energy (LSD) formula and Weizs\"acker-Skyrme (WS*) formula, are extracted through matching the energy per particle of finite nuclei. For LSD and WS*, the obtained incompressibility coefficients of symmetric nuclear matter are $K_\infty=230 \pm 11$ MeV and $235\pm 11$ MeV, respectively. The slope parameter of symmetry energy at saturation density is $L=41.6\pm 7.6$ MeV for LSD and $51.5\pm 9.6$ MeV for WS*, respectively, which is compatible with the liquid-drop analysis of Lattimer and Lim [ApJ. \textbf{771}, 51 (2013)]. The density dependence of the mean-field isoscalar and isovector effective mass, and the neutron-proton effective masses splitting for neutron matter are simultaneously investigated. The results are generally consistent with those from the Skyrme Hartree-Fock-Bogoliubov calculations and nucleon optical potentials, and the standard deviations are large and increase rapidly with density. A better constraint for the effective mass is helpful to reduce uncertainties of the depth of the mean-field potential.

\end{abstract}

\maketitle

\begin{center}
\textbf{I. INTRODUCTION}
\end{center}

Equation of state (EOS) for cold nuclear matter \cite{Myers98,Aich91}, e.g., the energy per particle of nuclear matter $e(\rho,\delta)=e(\rho,0)+E_{\rm sym}(\rho) \delta^2 + \mathcal{O}(\delta^4)$ considered as a function of the nuclear density $\rho$ and the isospin asymmetry $\delta=(\rho_n-\rho_p)/(\rho_n+\rho_p)$ where $\rho_n$ and $\rho_p$ denote neutron and proton densities, respectively, plays a key role in the interpretation of nuclear structure and nucleus-nucleus collisions, and as well as of neutron stars and supernova explosions. Its knowledge is therefore highly desirable. In addition to the properties of symmetric nuclear matter, especially the behavior of its density dependence \cite{Latt13,Li08,Chen05,Shet07,Botvina02,Stein,Stein05,Dong11,Khan12}, has also attracted a lot attention in recent years. The information of the symmetry energy at saturation and sub-saturation densities are obtained from nuclear dynamical behavior in heavy-ion collisions at intermediate and low energies \cite{Zhang08,Tsang09,Trip08}, and the static properties of finite nuclei such as neutron skin thickness \cite{Cent09,Wang13,Radii,Chen13} and nuclear masses \cite{Wang10,Wang14,HFB17,HFB27,Zhao10,Liu2010,Jiang12,Jiang15}. Although a great effort has been devoted in recent decades to investigate the properties of nuclear matter, the uncertainty of nuclear symmetry energy $E_{\rm sym}(\rho)$ is still large, for example, the slope parameter $L$ of the symmetry energy at the saturation density extracted from some independent analyses of various experimental observations are distributed in a range of $20 < L < 120$ MeV \cite{Zuo14}. Therefore, more investigations with high precision are still required.

As one of the basic quantities in nuclear physics, the nuclear masses can provide important information on the EOS at sub-saturation and saturation densities. For example, the energy per particle of symmetric nuclear matter and symmetry energy at saturation density can be estimated by the coefficient of volume term and symmetry energy coefficient in the liquid drop formula, respectively. Some nuclear mass models such as the Skyrme Hartree-Fock-Bogoliubov (HFB) models \cite{HFB17,HFB27} and the macroscopic-microscopic mass models \cite{Pom03,Wang10,Wang14}, have been successfully established with an rms error of $300 \sim 600$ keV with respect to more than 2000 measured nuclear masses. As macroscopic-microscopic mass formulas, both the Lublin-Strasbourg-Drop (LSD) formula \cite{Pom03} and the Weizs\"acker-Skyrme (WS*) formula \cite{Wang10,Wang14} use the Strutinsky's shell correction method for the microscopic part and similar liquid drop formula for the macroscopic energy of a spherical nuclei. Taking into account the curvature term in the liquid drop energy, the LSD formula can reproduce the masses in the latest nuclear mass datasets AME2012 \cite{Audi12} with an rms error of 608 keV \cite{Sob13}. Without taking into account the curvature term but considering the isospin dependence of model parameters, the WS* formula can reproduce the 2353 measured masses in AME2012 with an rms error of 439 keV \cite{Mo14}. It is known that nuclear masses of bound nuclei are significantly influenced by the behavior of $e(\rho,\delta)$ at around $\rho= 0.08 \sim 0.16$ fm$^{-3}$ and $|\delta|<0.4$, considering nuclear surface diffuseness \cite{Khan12}. Although the macroscopic-microscopic approaches are found to be the most accurate ones in the description of nuclear masses \cite{Sob13}, the information on the density dependence of energy per particle can not be directly obtained. One interesting question is how to extract the properties of cold neutron-rich nuclear matter at sub-saturation densities represented in the macroscopic-microscopic mass formulas.

It is known that the density functional theory is widely used in the study of the nuclear ground state which provides us with a useful balance between accuracy and computation cost, allowing large systems with a simple self-consistent manner. In the semi-classical ETF approach, the macroscopic energy of a nucleus can be self-consistently obtained by a given Skyrme energy density functional (EDF). When the energies per particle of a great number of stable and unstable nuclei predicted in the macroscopic-microscopic formulas can be remarkably well matched by the Skyrme EDF associated with a certain set of model parameters, one might indirectly obtain the properties of neutron-rich nuclear matter at densities around $\rho= 0.08 \sim 0.16$ fm$^{-3}$ by using the corresponding Skyrme EDF. It is also interesting to compare the Skyrme forces constrained from the macroscopic-microscopic mass formulas and those from Hartree-Fock calculations, since the treatment of microscopic effects is different.

In addition, considering the complexity of the parameter space in the Skyrme forces, the investigation of the uncertainty of model parameters is therefore important and necessary for assessing the model reliability and doing some extrapolations \cite{Roc15}. In this work, we will firstly match the liquid drop formula adopted in the LSD and WS* formulas by using the standard Skyrme EDF. Based on the obtained Skyrme forces, the corresponding density dependence of energy per particle and effective mass for symmetric and asymmetric nuclear matter will be investigated in the mean-field framework. Simultaneously, the standard deviations of some predicted quantities due to the uncertainty of matching procedure in the parameter space of Skyrme forces will be presented.

\newpage

\begin{center}
\textbf{II. MATCHING PROCEDURE}
\end{center}

According to the LSD mass formula, the ground state energy of a nucleus is expressed as a function of mass number $A$ and isospin asymmetry $I=(N-Z)/A$,
\begin{eqnarray}
E_{\rm LD}(A,I) \approx e_0 A + a_{\rm sym} I^2 A + ...,
\end{eqnarray}
neglecting the Coulomb energy, the Wigner energy (also called congruence energy) and the microscopic shell and pairing corrections. The binding energy per particle of a symmetric nucleus is expressed as,
\begin{eqnarray}
e_0(A)= a_v+a_s A^{-1/3}+ a_{\rm curv} A^{-2/3},
\end{eqnarray}
including the volume, surface and curvature terms.
The symmetry energy coefficient $a_{\rm sym}$ of a finite nucleus is written as
\begin{eqnarray}
a_{\rm sym}(A)=J-a_{\rm ss}A^{-1/3}+a_{\rm cs} A^{-2/3}
\end{eqnarray}
by using the Leptodermous expansion in terms of powers of $A^{-1/3}$. $J$ denotes the symmetry energy of nuclear matter at normal density. $a_{\rm ss}$ and $a_{\rm cs}$ are the coefficients of the surface-symmetry energy and curvature-symmetry energy terms, respectively.  The parameters of the liquid drop formula adopted in LSD and WS* are listed in Table I.

\begin{table} [h]
\centering
\caption{Model parameters of the LSD and WS* mass formulas (in MeV). }
\begin{tabular}{ccccccc}
 \hline\hline
Model & ~~~~$a_v$~~~~ & ~~~~ $a_s$ ~~~~ & ~~~~ $a_{\rm curv}$ ~~~~ & ~~~~ $J$ ~~~~ & ~~~~$a_{\rm ss}$~~~~ & ~~~~$a_{\rm cs}$~~~      \\
\hline
LSD	&	$-15.4920$ 	&	16.9707 	&	3.8602 	&	28.82 	&	38.93  &  	9.17  	     \\
WS*	&	$-15.6223$ 	&	18.0571 	&	$-$ 	&	29.16 	&	39.31  &    $-$	  	 	 \\
\hline
\end{tabular}
\end{table}

On the other hand, under the semi-classical ETF approximation \cite{Bencheikh, M.Brack,liumin}, the ``macroscopic" part of the nuclear energy of a nucleus can be expressed as an integral of the standard 10-parameter Skyrme EDF $\mathcal H({\bf r})$,
 \begin{eqnarray}
 \tilde{E} = \int {\mathcal H} [\rho_n({\bf r}), \rho_p({\bf r}))] \; d{\bf r},
\end{eqnarray}
since the kinetic energy density and the spin-orbit energy density can be expressed as a functional of nuclear density and its gradients.
Adopting the Fermi function
\begin{eqnarray}
\rho_q ({\bf r} )= \frac{\rho^{(q)}_{0}}{1+  \exp (\frac{  {\bf r} -R_{q} }{a_q})},
\end{eqnarray}
for describing the density distribution of a spherical nucleus ($q=n$ for neutrons and $q=p$ for protons), one can self-consistently obtain the minimal ``macroscopic" energy $\tilde{E}$ of the nucleus, by varying the four variables $R_{p}$, $a_{p}$, $R_{n}$, $a_{n}$ in Eq.(5) for a given nucleus. Here, $R_q$ and $a_{q}$ denote the radius and surface diffuseness of nuclei, respectively. The central density $\rho^{(q)}_{0}$ is determined from the conservation of particle number.

To match the nuclear liquid drop energy $E_{\rm LD}$ in the LSD formula for a series of finite nuclei with the corresponding ``macroscopic" energy $\tilde{E}$ from the Skyrme EDF, one could find the best-fit functional for the LSD mass formula. More specifically, adopting a certain set of Skyrme parameters in the literature as the initial values, we vary the 10 parameters ($t_0$, $t_1$, $t_2$, $t_3$, $x_0$, $x_1$, $x_2$, $x_3$, $\sigma$ and $W_0$) of the standard Skyrme EDF one by one in the 10-dimensional parameter space and calculate the corresponding $\tilde{E}$, and then search for the minimal rms deviation with respect to the LSD liquid drop energy $E_{\rm LD}$ by using downhill optimization algorithm \cite{liumin}. The calculations have been carried out not only for intermediate mass nuclei, but also for nuclei with huge numbers of nucleons, of the order of $10^6$, in order to perform a reliable extrapolation to neutron-rich nuclear matter. We search for the minimal rms deviation $ \sigma^2=  \frac{1}{m}\sum [\tilde{\varepsilon}^{(i)}-\varepsilon_{\rm LD}^{(i)}]^2 $ between $\tilde{\varepsilon}=\tilde{E}/A$ from the Skyrme force and $\varepsilon_{\rm LD}=E_{\rm LD}/A$ from the LSD formula for $m=70$ nuclei, with mass number $A=40, 60, 80, 100, 140, 200, 1000, 10^4, 10^5, 10^6$ and isospin asymmetry $I=0, 0.05, 0.1, 0.15, 0.2, 0.25, 0.3$. The Coulomb interaction has been ignored to be able to approach nuclei of arbitrary sizes and to avoid radial instabilities characteristic of systems with very large atomic numbers, as the same as those did in Ref. \cite{Rein06}.

\begin{center}
\textbf{III. RESULTS AND DISCUSSIONS}
\end{center}

\begin{figure}
\includegraphics[angle=-0,width= 0.75 \textwidth]{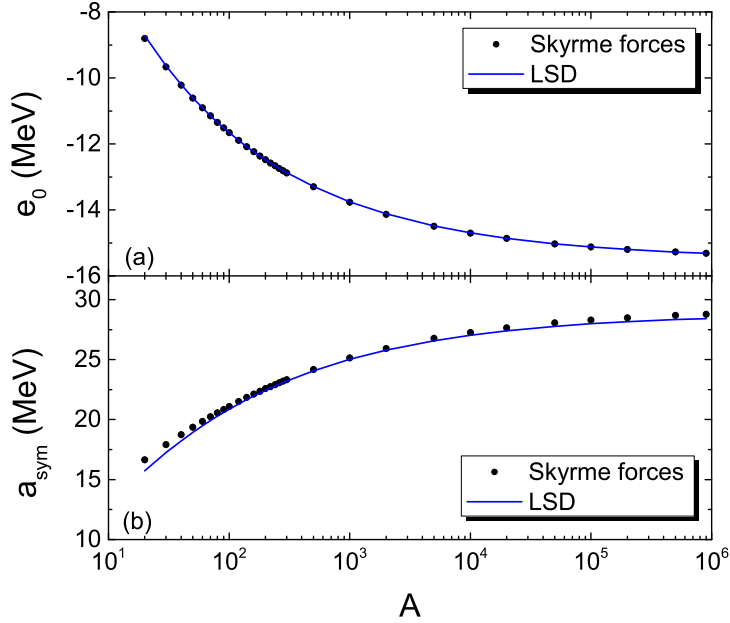}
 \caption{(Color online) Energy per particle (a) and symmetry energy coefficient (b) as a function of mass number.  The circles denote the corresponding mean values from the 82 Skyrme forces, and the standard deviations are smaller than the size of the symbols. The curves denote the results of the LSD formula according to Eq.(2) and (3).}
\end{figure}

Because of the complexity of the parameter space in the Skyrme forces, there exist probably many different Skyrme parameter sets leading to the similar rms deviations. To obtain the best fit functional from these similar rms deviations and to analyze the model uncertainty, we use 100 different Skyrme parameter sets in the literature (with the incompressibility of symmetric nuclear matter $K_\infty =235 \pm 35$ MeV) as the initial values. With a fit of the LSD liquid drop formula, we find 82 sets of new Skyrme parameters with which the minimal rms deviation between $\tilde{E}/A$ and $E_{\rm LD}/A$ is only $\sigma=8\pm 2$ keV for the 70 nuclei. Based on the same approach proposed in Ref. \cite{Wang15}, we extract the mass dependence of $e_0(A)$ and of symmetry energy coefficient $a_{\rm sym}(A)$ for the obtained 82 Skyrme parameter sets. The corresponding results are shown in Fig. 1. The curves in the figure denote the corresponding results of the LSD model according to Eq.(2) and (3). One sees that both $e_0(A)$ and $a_{\rm sym}(A)$ in the LSD formula can be remarkably well reproduced with the obtained Skyrme forces except very light nuclei.

\begin{figure}
\includegraphics[angle=-0,width= 0.8 \textwidth]{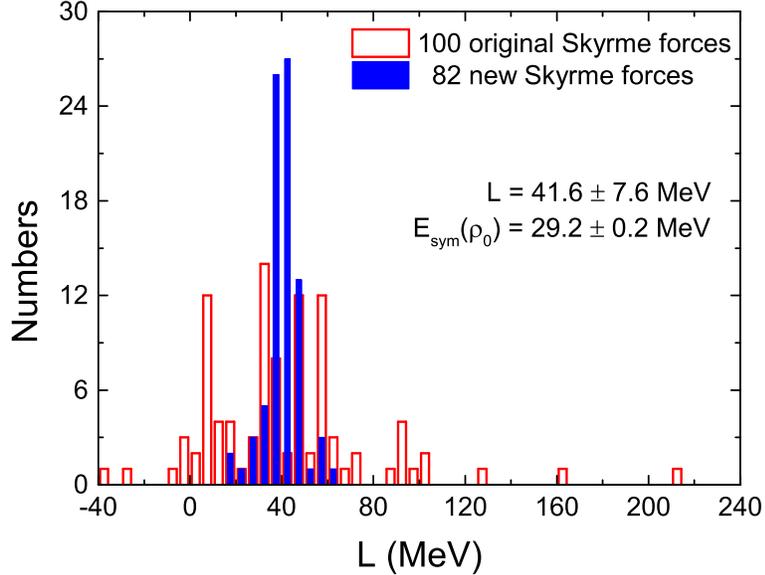}
 \caption{(Color online) Distribution of the density slope of $E_{\rm sym}$ at normal density. The red hollow bars denote the results from 100 original Skyrme parameter sets in the literature. The blue solid bars denote the results of 82 new Skyrme forces after matching the LSD formula, with mean value of $L=41.6$ MeV and standard deviation of $7.6$ MeV. }
\end{figure}

With the obtained 82 new Skyrme forces for the LSD formula, the properties of nuclear matter can be further investigated.
For nuclear matter, the symmetry energy in the standard Skyrme EDF is expressed as
\begin{eqnarray}
E_{\rm sym}(\rho) &=& \frac{1}{3}\frac{\hbar^2}{2m} \left ( \frac{3\pi^2}{2}
\right )^{2/3} \rho^{2/3}-  \frac{1}{8} t_0 (2 x_0+1)\rho  \nonumber \\
&-&   \frac{1}{24} \left(
 \frac{3 \pi^2}{2} \right )^{2/3} \Theta_{\rm sym}
 \rho^{5/3}-\frac{1}{48} t_3 (2x_3+1) \rho^{\sigma+1}
\end{eqnarray}
with $ \Theta_{\rm sym}=3 t_1 x_1 -t_2(4+5 x_2)$.
The slope parameter of the symmetry energy at normal density $\rho_0$ is written as,
\begin{eqnarray}
L =3\rho_0 \left (\frac{\partial E_{\rm sym}}{\partial \rho} \right
)_{\rho=\rho_0}.
\end{eqnarray}
Fig. 2 shows the distribution of the values of $L$ calculated from these different Skyrme forces. The red hollow bars denote the results of the 100 sets of Skyrme forces in the literature, and the blue solid bars denote the corresponding results of the obtained 82 Skyrme forces after matching the LSD liquid drop energy. The values of $L$ from the considered Skyrme forces in the literature distribute in a very large range, from about $-40$ MeV to 210 MeV. After matching the LSD formula, we find the values of $L$ focus on a small region with the mean value of $41.6$ MeV and the standard deviation of $7.6$ MeV. The corresponding symmetry energy at saturation density is $E_{\rm sym} (\rho_0)=29.2\pm 0.2$ MeV based on the obtained 82 Skyrme parameter sets.

With the same approach, we also study the WS* mass formula. In the WS* formula, the curvature terms are not considered. We find that the liquid drop energy $E_{\rm LD}$ in the WS* formula is not matched as good as that in the LSD formula, probably due to the influence of the curvature terms. We obtain 74 new Skyrme parameter sets in which the minimal rms deviations with respect to $E_{\rm LD}/A$ of the 70 nuclei are $\sigma=30\pm 5$  keV.  The corresponding quantities related to the equation of state according to the obtained Skyrme forces are listed in Table II. $e_\infty=\frac{E}{A}(\rho_0)$ and $K_\infty$ denote the energy per particle of symmetric nuclear matter and its curvature at the saturation density $\rho_0$, respectively. $K_{\rm sym}$ denotes the curvature of the symmetry energy at $\rho_0$. $E_{\rm sym} (\rho_c)$ and $L_c$ denote the symmetry energy and its slope at sub-normal density of $\rho_c=0.1$ fm$^{-3}$, respectively. By adopting different Skyrme parameter sets, one can obtain the distribution of a certain quantity (see Fig. 2 for example), and consequently the mean value and the corresponding standard deviation can be obtained. In this work, the uncertainty of model predictions for a quantity is described by its standard deviation. Here, we would like to state that the uncertainty in this work is due to the uncertainty in the fit of the mass formulas from the parameter space of Skyrme forces, rather than directly from the experimental observations. From the table, one sees that the value of $L$ for WS* is larger than that for LSD by about 10 MeV. At sub-normal density of $\rho_c=0.1$ fm$^{-3}$, the discrepancy between the corresponding slope parameters $L_c$ falls to $\sim 4.4$ MeV, and the symmetry energy $E_{\rm sym} (\rho_c)$ of these two formulas is very close to each other. The extracted slope parameter $L= 54 \pm 19$ MeV from the charge radii of $^{30}$S - $^{30}$Si mirror pair \cite{Radii}, $L=52.5 \pm 20$ MeV from the Skyrme Hartree-Fock calculations together with the neutron skin thickness of Sn isotopes \cite{Chen11} and $L=52.7 \pm 22.5$ MeV from the global nucleon optical potentials \cite{Xu10} are in good agreement with the estimated result for the WS* formula. In addition, the symmetry energy and its slope parameter obtained for the two mass models are compatible with the liquid-drop analysis of Lattimer and Lim \cite{Latt13}.

 \begin{table}
\centering
\caption{Quantities related to EOS matched for the LSD and WS* formulas (in MeV). }
\begin{tabular}{cccccccc}
 \hline\hline
Model & ~~~~~~$e_\infty$~~~~~~~~~~~~~~ &~~$K_\infty$~~ & ~~ $E_{\rm sym} (\rho_0)$ ~~ & ~~ $L$ ~~ &~~ $K_{\rm sym}$ ~~ & ~~$L_c$~~& ~~ $E_{\rm sym} (\rho_c)$ ~~    \\
\hline
LSD	&	$-15.494\pm 0.004$ &$230\pm 11$ 	&	$29.2 \pm 0.2$ 	&	$41.6\pm 7.6$ 	&	$-152 \pm 41$ 	&	$40.4\pm 2.4$  &	$23.1 \pm 0.5$   	     \\
WS*	&	$-15.583\pm 0.007$ &$235\pm 11$ 	&	$29.7 \pm 0.3$ 	&	$51.5\pm 9.6$ 	&	$-117 \pm 46$ 	&	$44.8\pm 3.4$  &	$23.3 \pm 0.7$ 	 	 \\
\hline
\end{tabular}
\end{table}

\begin{figure}
\includegraphics[angle=-0,width= 0.9 \textwidth]{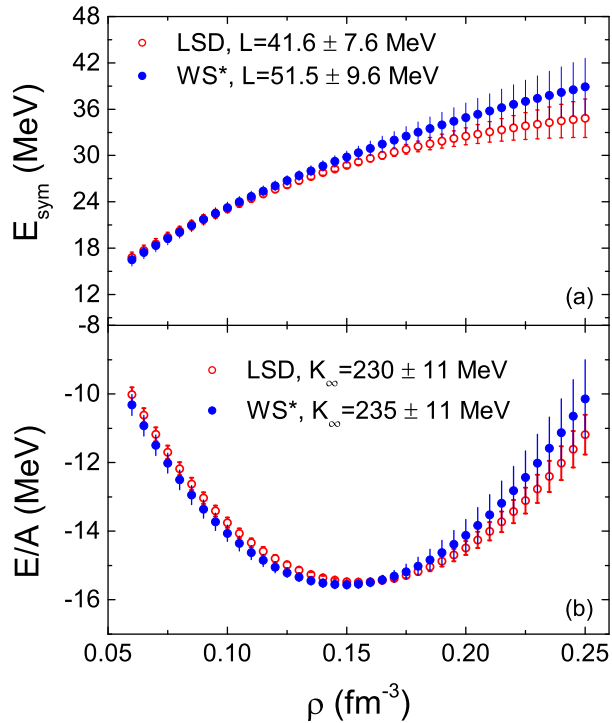}
 \caption{(Color online)  Density dependence of nuclear symmetry energy (a) and energy per particle for symmetric nuclear matter (b). The open and solid circles denote the results by matching the LSD and WS* formulas, respectively.}
\end{figure}

In Fig. 3 (a), we show the density dependence of symmetry energy with the best-fit parameter sets for the LSD and WS* formulas. The error bars denote the corresponding standard deviations.  One sees that at sub-normal density region, the symmetry energy from these two models is very close to each other. Whereas at the region $\rho>\rho_0$, the standard deviation increases rapidly with the increase of density, which indicates that only with nuclear masses the nuclear symmetry energy at supra-saturation densities can not be accurately constrained. In Fig. 3(b), we show the obtained energy per particle of symmetric nuclear matter for the LSD and WS* formulas as a function of density. The obtained incompressibility coefficient $K_\infty$ for the WS* formula is comparable with that for the LSD formula, and the values for both models are in good agreement with the generally accepted value of $K_\infty \approx 230 $ MeV \cite{Cent09,Khan12} .

Together with the symmetry energy, the splitting of neutron and proton effective masses in neutron-rich matter is also an important quantity related to the isospin-dependence of nucleon-nucleon interaction. Whether the effective mass $m^*_n$ for neutrons is higher than that $m^*_p$ for protons in neutron-rich matter or the magnitude of effective mass splitting changes as a function of momentum \cite{LongWH} is an interesting question and attracted a lot of attention in recent years \cite{ZhangYX,ZhangYX13,Li15}. Here, based on the obtained Skyrme forces for the LSD and WS* formulas we simultaneously investigate the density dependence of the mean-field isoscalar effective mass and the splitting of neutron and proton effective masses. In the framework of Skyrme EDF, the isoscalar and isovector effective masses are written as \cite{Lesi06},
\begin{eqnarray}
\frac{m^*_s}{m}=\frac{1}{1+\kappa_s },
\end{eqnarray}
with $\kappa_s=\frac{2m}{\hbar^2} \frac{1}{16}[3t_1+t_2(5+4x_2)] \rho$ and
\begin{eqnarray}
\frac{m^*_v}{m}=\frac{1}{1+\kappa_v },
\end{eqnarray}
with $\kappa_v=\frac{2m}{\hbar^2} \frac{1}{8}[2(t_1+ t_2)-t_1 x_1+t_2 x_2] \rho$, respectively.
The splitting of neutron and proton effective masses for neutron matter ($\delta=1$) is expressed as \cite{Lesi06},
\begin{eqnarray}
\frac{\Delta m^*}{m}=\frac{m^*_{n}-m^*_{p}}{m}=\frac{2(\kappa_v-\kappa_s)}{(1+\kappa_s)^2-(\kappa_v-\kappa_s)^2}.
\end{eqnarray}

 \begin{figure}
\includegraphics[angle=-0,width= 0.8 \textwidth]{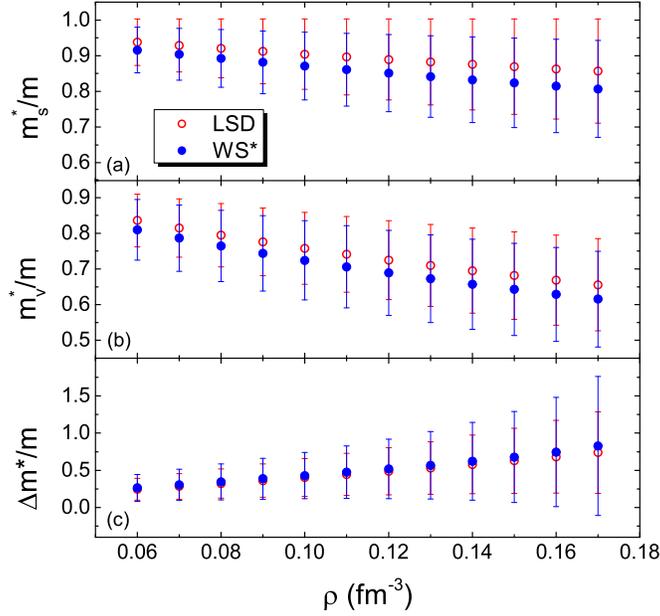}
 \caption{(Color online) (a) Mean-field isoscalar effective mass, (b) isosvector effective mass, and (c) splitting of neutron and proton
effective masses as a function of density.}
\end{figure}

Fig. 4 shows the calculated $\frac{m^*_s}{m}$, $\frac{m^*_v}{m}$ and $\frac{\Delta m^*}{m}$ as a function of density. With the increase of density, the isoscalar effective mass decreases linearly at sub-saturation density region in general. At saturation density, the mean values of the extracted isoscalar and isovector effective masses for LSD are about 0.86 and 0.67, respectively. For WS*, $\frac{m^*_s}{m}\approx 0.82$ and $\frac{m^*_v}{m}\approx 0.63$ at saturation density, which are roughly comparable with the corresponding values (0.8 and 0.72) given in the HFB-27 model \cite{HFB27}. For the splitting of neutron and proton effective masses of neutron matter, the obtained results for the two models are similar, with positive values and increasing with the density. The neutron effective mass $m^*_n$ is larger than the proton effective mass $m^*_p$ in neutron-rich matter, which is consistent with measurements of isovector giant resonances \cite{Lesi06}, the Skyrme HFB calculations \cite{HFB17} in general. Very recently, Xiao-Hua Li et al. investigated the neutron-proton effective mass splitting from the global nucleon optical potentials \cite{Li15}. The estimation of  $\frac{\Delta m^*}{m}=0.41 \pm 0.15$ for neutron matter around normal density is also generally consistent with the results in this work. In addition, one can see that the corresponding standard deviations in Fig. 4 are large and increase rapidly with the density.

 \begin{figure}
\includegraphics[angle=-0,width= 0.7 \textwidth]{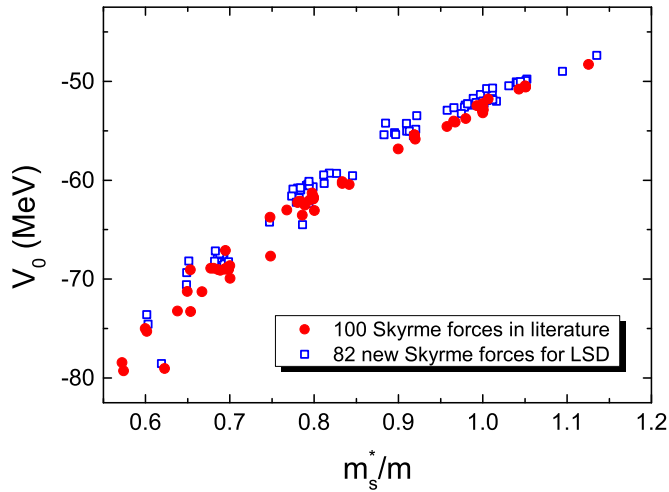}
 \caption{(Color online) Depth of single particle potential for symmetric nuclear matter at the saturation density as a function of isoscalar effective mass. }
\end{figure}

In this work, we also investigate the depth of the single-particle potential from the considered Skyrme forces and its correlation with the effective mass. It is thought that the effective mass is related to the depth of the single particle potential \cite{Zuo14,Li15}. Fig. 5 shows the calculated depth of the mean-field potential for symmetric nuclear matter at the saturation density as a function of isoscalar effective mass based on the Skyrme forces considered. One sees that either from the 100 Skyrme forces in the literature (circles) or from the ones for LSD (squares), the depth of the potential evidently increases with the corresponding isoscalar effective mass as expected. The potential depth from the zero-range Skyrme forces considered in Fig. 5 varies in the region of $V_0 = 63 \pm 17 $ MeV, whereas the depth of phenomenological Woods-Saxon single particle potential is usually about $V_0 \approx50 $ MeV \cite{Cwoik,Wang10}. It is known that the single particle picture is valid mostly around the Fermi energy and it is unrealistic to directly measure the depth of single particle potential. Comparing with the difficulties in the measurement of the depth of the single particle potential, it is much more realistic to measure the value of effective mass, from level densities, collective modes, etc. The correlation between the potential depth and the effective mass indicates that a better determination of the effective mass would be helpful to reduce the uncertainties of the depth of the mean-field potential.

\begin{center}
\textbf{IV. SUMMARY}
\end{center}

By using the extended Thomas-Fermi approximation together with the restricted density variational method, the corresponding properties for symmetric and asymmetric nuclear matter represented in the LSD and WS* mass formulas has been investigated with the standard 10-parameter Skyrme energy density functionals. Through matching the nuclear liquid drop energy given in the LSD and WS* formulas for finite nuclei with the corresponding ``macroscopic" energy calculated from the Skyrme energy density functionals, we attempt to obtain the information on the density dependence of binding energy and symmetry energy represented in the macroscopic-microscopic mass models. We find that LSD liquid drop formula can be remarkably well reproduced by 82 new Skyrme forces after adjusting the Skyrme parameters, with an rms error of only about 8 keV. For the WS* formula, the liquid drop energy is not matched as good as that of LSD due to the neglecting of the curvature terms. The obtained slope parameter of symmetry energy is $L=41.6\pm 7.6$ MeV for LSD and $51.5\pm 9.6$ MeV for WS*. The predicted symmetry energies at sub-saturation density region from these two mass formulas are very close to each other. At supra-saturation density region, the uncertainties of the energy per particle and symmetry energy increase rapidly with the density.

Based on the new Skyrme forces for LSD and WS*, the correlation between the symmetry energy and its slope parameter is also observed evidently. The slope parameter $L$ generally increases with the symmetry energy $J$. In the present work, the two mass models have a relatively low value for the symmetry energy, around 29 MeV, and therefore the corresponding slope parameters have relatively low values comparing with other predictions, such as the relativistic calculations. One should note that the extracted information in this work is still model dependent, considering the limitation of the non-relativistic standard Skyrme energy density functional and the correlation between symmetry energy and its slope parameter.

With the obtained Skyrme forces for the macroscopic-microscopic formulas, the density dependence of the mean-field isoscalar and isovector effective mass and the splitting of neutron and proton effective masses are simultaneously investigated. The results are generally consistent with those from the Skyrme HFB calculations and nucleon optical potentials around saturation density. The large standard deviations for the effective mass from nuclear mass models imply that other constraints are still required to obtain the information on the behavior of effective mass. Considering the correlation between the potential depth and the effective mass, a better constraint for the effective mass is helpful to reduce uncertainties of the depth of the mean-field potential.

\begin{center}
\textbf{ACKNOWLEDGEMENTS}
\end{center}

We thank Lie-Wen Chen for reading of the manuscript and helpful communications, and an anonymous referee for valuable comments.  This work was supported by National Natural Science Foundation of China (Nos 11275052, 11365004, 11365005, 11475262, 11422548, 11265004), National Key Basic Rsearch Development Program of China under Grant No. 2013CB834404, and the Open Project Program of State Key Laboratory of Theoretical Physics, Institute of Theoretical Physics, Chinese Academy of Sciences, China (No. Y4KF041CJ1).

\end{document}